\documentstyle{mn}

\newif\ifAMStwofonts

\input epsf
\newcommand{\etal}{et al.}
\newcommand{\etals}{\etal \ }
\newcommand{\um}{$\mu m$}
\newcommand{\ums}{\um \ }
\newcommand{\Lsol}{L$_\odot$}
\newcommand{\Msol}{M$_\odot$}
\newcommand{\Msols}{\Msol \ }

\newcommand{\figs}[3]{
\begin{figure}
\vspace{0cm}
\hspace{0cm}\epsfxsize=8.8cm \epsfbox{#1}
\vspace{0cm}
\caption{#2}
\label{#3}
\end{figure}
}

\ifoldfss
  \ifCUPmtlplainloaded \else
    \NewTextAlphabet{textbfit} {cmbxti10} {}
    \NewTextAlphabet{textbfss} {cmssbx10} {}
    \NewMathAlphabet{mathbfit} {cmbxti10} {} 
    \NewMathAlphabet{mathbfss} {cmssbx10} {} 
  \fi
  \ifAMStwofonts
    \ifCUPmtlplainloaded \else
      \NewSymbolFont{upmath} {eurm10}
      \NewSymbolFont{AMSa} {msam10}
      \NewMathSymbol{\upi}     {0}{upmath}{19}
      \NewMathSymbol{\umu}     {0}{upmath}{16}
      \NewMathSymbol{\upartial}{0}{upmath}{40}
      \NewMathSymbol{\leqslant}{3}{AMSa}{36}
      \NewMathSymbol{\geqslant}{3}{AMSa}{3E}

    \fi
  \fi
\fi 

\ifnfssone
  \newmathalphabet{\mathit}
  \addtoversion{normal}{\mathit}{cmr}{m}{it}
  \addtoversion{bold}{\mathit}{cmr}{bx}{it}
  \newmathalphabet{\mathbfit} 
  \addtoversion{normal}{\mathbfit}{cmr}{bx}{it}
  \addtoversion{bold}{\mathbfit}{cmr}{bx}{it}
  \newmathalphabet{\mathbfss} 
  \addtoversion{normal}{\mathbfss}{cmss}{bx}{n}
  \addtoversion{bold}{\mathbfss}{cmss}{bx}{n}
  \ifAMStwofonts
    \ifCUPmtlplainloaded \else
      \UseAMStwoboldmath
      \makeatletter
      \new@mathgroup\upmath@group
      \define@mathgroup\mv@normal\upmath@group{eur}{m}{n}
      \define@mathgroup\mv@bold\upmath@group{eur}{b}{n}
      \edef\UPM{\hexnumber\upmath@group}
      \new@mathgroup\amsa@group
      \define@mathgroup\mv@normal\amsa@group{msa}{m}{n}
      \define@mathgroup\mv@bold\amsa@group{msa}{m}{n}
      \edef\AMSa{\hexnumber\amsa@group}
      \makeatother
      \mathchardef\upi="0\UPM19
      \mathchardef\umu="0\UPM16
      \mathchardef\upartial="0\UPM40
      \mathchardef\leqslant="3\AMSa36
      \mathchardef\geqslant="3\AMSa3E
    \fi
  \fi
\fi 

\ifnfsstwo
  \DeclareMathAlphabet{\mathbfit}{OT1}{cmr}{bx}{it}
  \SetMathAlphabet\mathbfit{bold}{OT1}{cmr}{bx}{it}
  \DeclareMathAlphabet{\mathbfss}{OT1}{cmss}{bx}{n}
  \SetMathAlphabet\mathbfss{bold}{OT1}{cmss}{bx}{n}
  \ifAMStwofonts
    \ifCUPmtlplainloaded \else
      \DeclareSymbolFont{UPM}{U}{eur}{m}{n}
      \SetSymbolFont{UPM}{bold}{U}{eur}{b}{n}
      \DeclareSymbolFont{AMSa}{U}{msa}{m}{n}
      \DeclareMathSymbol{\upi}{0}{UPM}{"19}
      \DeclareMathSymbol{\umu}{0}{UPM}{"16}
      \DeclareMathSymbol{\upartial}{0}{UPM}{"40}
      \DeclareMathSymbol{\leqslant}{3}{AMSa}{"36}
      \DeclareMathSymbol{\geqslant}{3}{AMSa}{"3E}
    \fi
  \fi
\fi 

\ifCUPmtlplainloaded \else
  \ifAMStwofonts \else 
    \def\upi{\pi}
    \def\umu{\mu}
    \def\upartial{\partial}
  \fi
\fi

\title[ISO Observations of BR1202-0725]
  {ISO Observations of the dusty quasar BR1202-0725}
\author[K.J. Leech et al.]
  {K.J. Leech,$^1$ 
  L. Metcalfe,$^1$ B. Altieri,$^2$\\
  $^1$ISO Data Centre, Astrophysics Division, Space Science Department of
   ESA, Villafranca del Castillo, P.O. Box 50727, \\ 28080 Madrid, Spain \\
  $^2$XMM-Newton Data Centre, Astrophysics Division, Space Science Department 
   of ESA, Villafranca del Castillo, P.O. Box 50727, \\ 28080 Madrid, Spain}
\date{date}
\pagerange{\pageref{firstpage}--\pageref{lastpage}}
\pubyear{2000}

\begin{document}

\maketitle  



\begin{abstract}
We present mid- and far-IR photometry of the high-redshift (z=4.69) dusty 
quasar BR1202-0725. The quasar was detected in the near-IR, at a flux level 
(0.7$\pm$0.2 mJy) consistent with an average Radio-Quiet Quasar at it's 
redshift. Only upper limits for the emission were obtained in the far-IR. These 
upper limits, when combined with data from ground-based telescopes, are the 
first direct evidence for a turn-over in the far-IR emission and hence confirm 
that a black-body dominates the SED at FIR wavelengths. This black-body is most
probably cool dust, constrained to have a temperature below 80K, for a $\beta$ 
of 1.5.
\end{abstract}

\begin{keywords}
galaxies : active -- galaxies : photometry -- infrared : galaxies -- quasars :
individual : BR1202-0725
\end{keywords}

\section{Introduction}

It came as a surprise when groups working with IRAS data, e.g. Soifer \etals 
\shortcite{soi84} \& Lawrence \etals \shortcite{l86}, discovered galaxies
with bolometric luminosities equivalent to those of quasars but emitting
predominantly in the infra-red wavelength region. Even more surprising was 
the discovery by Rowan-Robinson \etals \shortcite{mrr91} of IRAS FSC 
10214+4724, a galaxy at a redshift of z=2.29 with an apparent IR luminosity 
of 4$\times$10$^{14}$ \Lsol. While F10214+4724 has turned out to be 
gravitationally lensed \cite{e96}, leading to a lower luminosity of 
2$\times$10$^{13}$ \Lsol, it is still of extreme luminosity.

Five high-redshift radio-quiet QSO's were selected, primarily from an
APM catalogue, by McMahon \etals \shortcite{mcm94} because their similarity to 
F10214+4724 in redshift and overall properties made them natural objects in 
which to look for IR-emission from dust. BR1202-0725 (see figure \ref{KL:fig1} 
for a DSS image of the quasar and surrounding field), at a redshift of 4.69, 
was the only one to be detected at 1.25 mm with IRAM. Assuming the detection 
was due to dust and that the dust had a single temperature (80K was assumed), 
McMahon \etals \shortcite{mcm94} derived a dust mass of the order of 
4$\times$10$^{8}$ \Msol, a considerable amount of dust to exist at such an 
early epoch. If the far-IR spectrum of BR1202-0725 is similar to that of 
F10214+4724 then it has an IR luminosity of the order of 10$^{14}$\Lsol, making 
it one of the most luminous objects in the universe.

Later IR observations of this object by Isaak \etals \shortcite{isa94} at 450, 
800 \& 1100 \um, and by Benford \etals \shortcite{ben99} at 350 microns, 
supported a slightly lower dust temperature of 68K and 50$\pm$7 K respectively, 
although none of these observations saw the turnover of the emission 
to confirm that it comes from a black-body. Indeed, its spectral index of 
2.7$\pm$0.4 between 1100 and 800\um, calculated by Isaak \etal, does not 
formally rule out synchrotron-emission, although a spectral index above 2.5 is 
difficult to explain by anything other than cool dust.

The detection of CO emission from approximately 1$\times$10$^{11}$\Msols of 
molecular hydrogen (comparable to that in a present-day luminous galaxy) by 
Ohta \etals \shortcite{oha96} and Omont \etals \shortcite{omo96b}, and the
detection by Omont \etals \shortcite{omo96b} of a close companion also 
containing large amounts of molecular gas, further reinforced the idea that 
the galaxy was undergoing a large burst of star formation, possibly interaction
triggered.

\figs{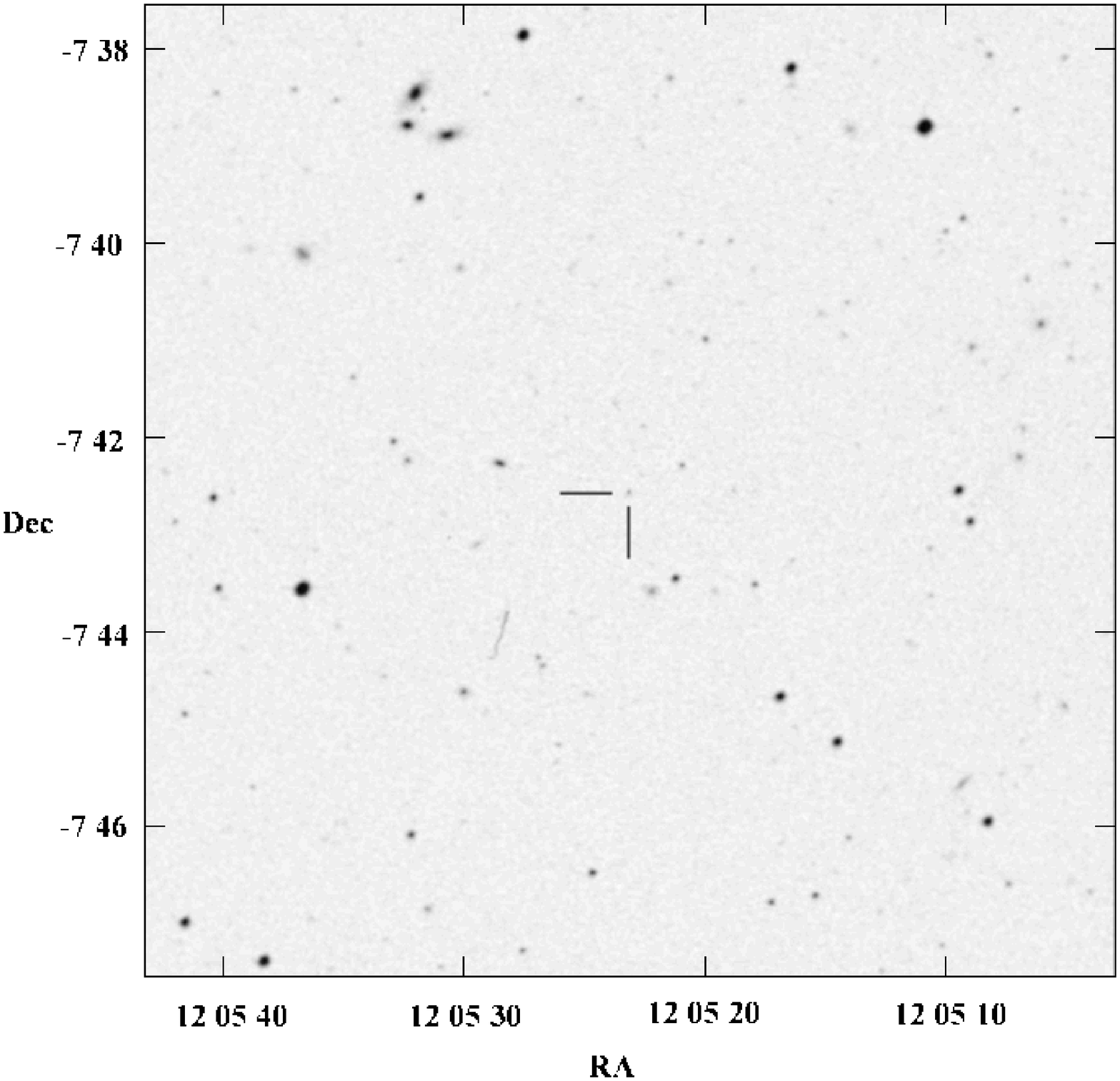}{The field around the quasar BR 1202-0725, taken 
from the DSS. The quasar is in the middle of the field and is indicated. The 
scale is consistent with figure \ref{fig:phot}.}{KL:fig1}

\section{Observations and data reduction}

\subsection{Observing Strategy}

The observations of BR1202-0725 had two goals: to try and detect the 
turnover in the far-IR emission; and to detect the near-IR emission from the 
continuum to better determine the underlying power-law. Early predictions from 
Isaak \etals \shortcite{isa94} were for flux densities in the 100 -- 200\ums 
range to reach of the order of 100 mJy, depending on dust temperature. We 
therefore proposed to make observations through the wideband C90 and C160 
filters (centred at 90 and 170 \um) of the ISOPHOT instrument \cite{lem96} 
on-board ESA's Infrared Space Observatory (ISO, Kessler \etals 1996). It was 
hoped these two far-IR observations would lie either side of the peak of the 
emission.

The ISOCAM instrument, described in Cesarsky \etals \shortcite{ces96}, was used
to obtain a single observation through the LW10 (8 -- 15 \ums) filter with the
aim of detecting the near-IR emission of BR1202-0725. A normal RQQ quasar with 
the redshift of BR1202-0725 would have a flux of approximately 1mJy in this 
band, and the ISOCAM observation was therefore planned to reach that 
sensitivity by employing the microscanning mode.

\subsection{Observations}

Observations, listed in Table \ref{KL:tab1}, were performed on July 14 1996. 
The ISOCAM observation was taken in microscanning mode, with the field-of-view 
of the instrument scanned over the object of interest in a 3$\times$3 raster 
pattern. The detector was read out every 2.1 seconds, leading to redundancy in 
the dataset. The redundancy is used to distinguish glitches and other 
artifacts from valid sources. The ISOPHOT observations were taken as PHT32 
rasters of size 3$\times$4. The ISOCAM \cite{sie99} and ISOPHOT \cite{lau00} 
Handbooks give complete descriptions of these observing modes.

\begin{table}
 \caption{ISO observations, taken on 14 July 1996.}
 \label{KL:tab1}
 \begin{tabular}{@{}lllll}
Observation     &TDT            &Wavelength     &Filter &Exposure \\
Mode            &number         &\um            &       &secs     \\
\hline
CAM01           &24002102       &8--15          &LW10   &348      \\
PHT32           &24002103       &90             &C90    &606      \\
PHT32           &24002101       &170            &C160   &2696     \\
\end{tabular}
\end{table}

\subsection{Data reduction}

\subsubsection{CAM Data Reduction}

The data was sliced and dark corrected in the standard way (Delaney 2000),
making use of a time-dependent dark correction. A thorough first-order
deglitching was performed using an iterative sigma-clipping method (Metcalfe et
al. in prep). 

No correction was made at this stage for the effects of responsive 
transients as the existing correction algorithms can adversely affect
the S/N for very faint sources. The transient correction was instead
made at the end of the data reduction by applying an appropriate
scaling factor. This factor was established by transient correcting
reference sources identified for this purpose in representative data
sets.

Dedicated faint-source processing followed essentially the method
described in Altieri \etals \shortcite{al98} and Metcalfe et al. (in 
prep).

Long-term baseline drifts were removed by smoothing (via median
filtering) the time-history of each individual detector pixel, as
sampled by the hundreds of readouts made during the observation, and
then subtracting the smoothed baseline from the nominal history. The 
images recorded at the several raster positions were organised into a
cube so that all samples of a given sky-position line-up. For each sky position,
this sky-position-vector is then sigma-clipped to give a second-order
deglitching which has been found to be extremely effective in
removing residual glitches. It should be noted that first-order
deglitching acts on the time-history of a detector pixel, as it moves
over the sky during a raster. Second-order deglitching acts on the
history of samples of each sky position, and so takes implicit
advantage of the sampling redundancy of the raster measurement
technique. The two deglitching steps are therefore highly
complementary. Following this deglitching stage the raster mosaic was
constructed straightforwardly from the raster-position-images and
standard aperture photometry could be performed on the resulting map.

In order to calibrate the aperture photometry the data were re-processed after 
inserting a set of fake sources into the raw data cube. These consisted of 
theoretical model PSFs matched to the optical configuration in use. By 
performing aperture photometry on these fake sources in a manner identical to 
that used for the actual source any flux-altering effects of the complex 
data-reduction algorithm could be calibrated.

The fake-source insertion operation was repeated for a range of
fake-source brightnesses. In this way the source strength in ADUgs (ADU
per gain per second) in the raw data, corresponding to a given source
strength in the reduced map, was determined.

At this point it was only necessary to scale the source strength in the
raw data for the effects of responsive transients (as described above)
and to apply the standard calibration scaling factor relating ADUgs to
incident mJy for each CAM filter (Siebenmorgen \etals 1999), in order to
arrive at an estimate of the flux of the target source.

\subsubsection{ISOPHOT Data Reduction}

The ISOPHOT P32 data was reduced using the PHOT Interactive Analysis (PIA) 
package including a pre-release version of P32 processing routines developed by 
R. Tuffs (Tuffs \etals in prep.). Initial data processing (ERD to SCP) used 
the new P32 processing routines, with the later processing (SCP to AAP) using 
the standard processing routines. Default values for the reduction 
parameters were used in both the old and new routines. 

Maps were generated using the Trigrid interpolation and the first
quartile normalization flat-fielding methods.

\section{Results}

In the ISOCAM dataset, see Figure \ref{fig:cam}, BR1202-0725 was detected with 
a flux of 0.7 mJy. Deriving the scatter in photometry of ten randomly placed 
apertures near the centre of the image produces a 1$\sigma$ value of 0.07 mJy, 
while the results of the simulations give a (random plus systematic) 
uncertainty of 0.2 mJy. We therefore take a value of 0.7$\pm$0.2 mJy for the 
flux from this object in the 8--15 \ums wavelength region. The background flux 
was determined to be 1.06$\pm$0.1 mJy per sq.arcsec (45$\pm$4 MJy sr$^{-1}$), 
consistent with the typical background predictions for CAM LW10 given in the 
user manual.

Other surrounding sources (visible on the DSS image) are also present in the 
ISOCAM image. As they are not in the centre of the image the observation was 
not optimised for them, leading to a higher uncertainty on their detections, 
and we therefore do not quote fluxes for them.

\figs{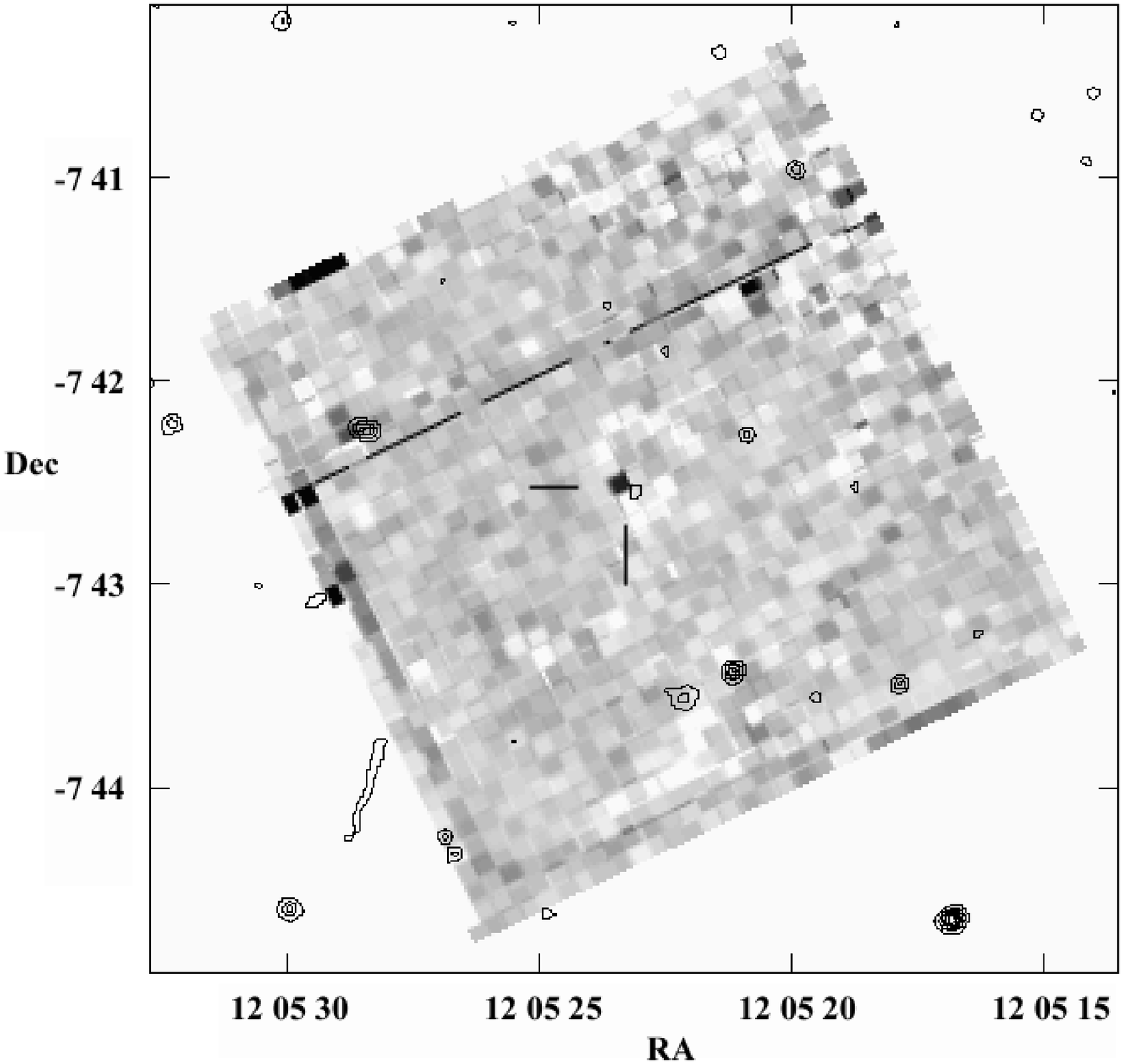}{The ISOCAM image overplotted with contours from a
DSS2 red image. The quasar is in the centre of the image, with the black line 
above being the missing column 24. There is a slight shift between the two
images of approximately 5$^{\prime\prime}$. This offset is not significant as 
the uncertainty in the ISOCAM astrometry due to the wheel positioning jitter 
is greater than this. The position of the quasar is marked.}{fig:cam}

No object was detected at the position of the quasar in either the 90\ums or 
170\ums datasets. The ISOPHOT 170\ums map is shown in figure \ref{fig:phot}. 
The effective noise in both maps was derived by moving a box the size of one 
C100 or C200 array pixel over the two maps (avoiding the outer parts of the
maps) and converting to Jy. This produced 1 sigma noise values of 16 and 15 mJy 
at 90 and 170\ums -- effectively the 3 sigma upper limit for both maps is 50 
mJy.

\figs{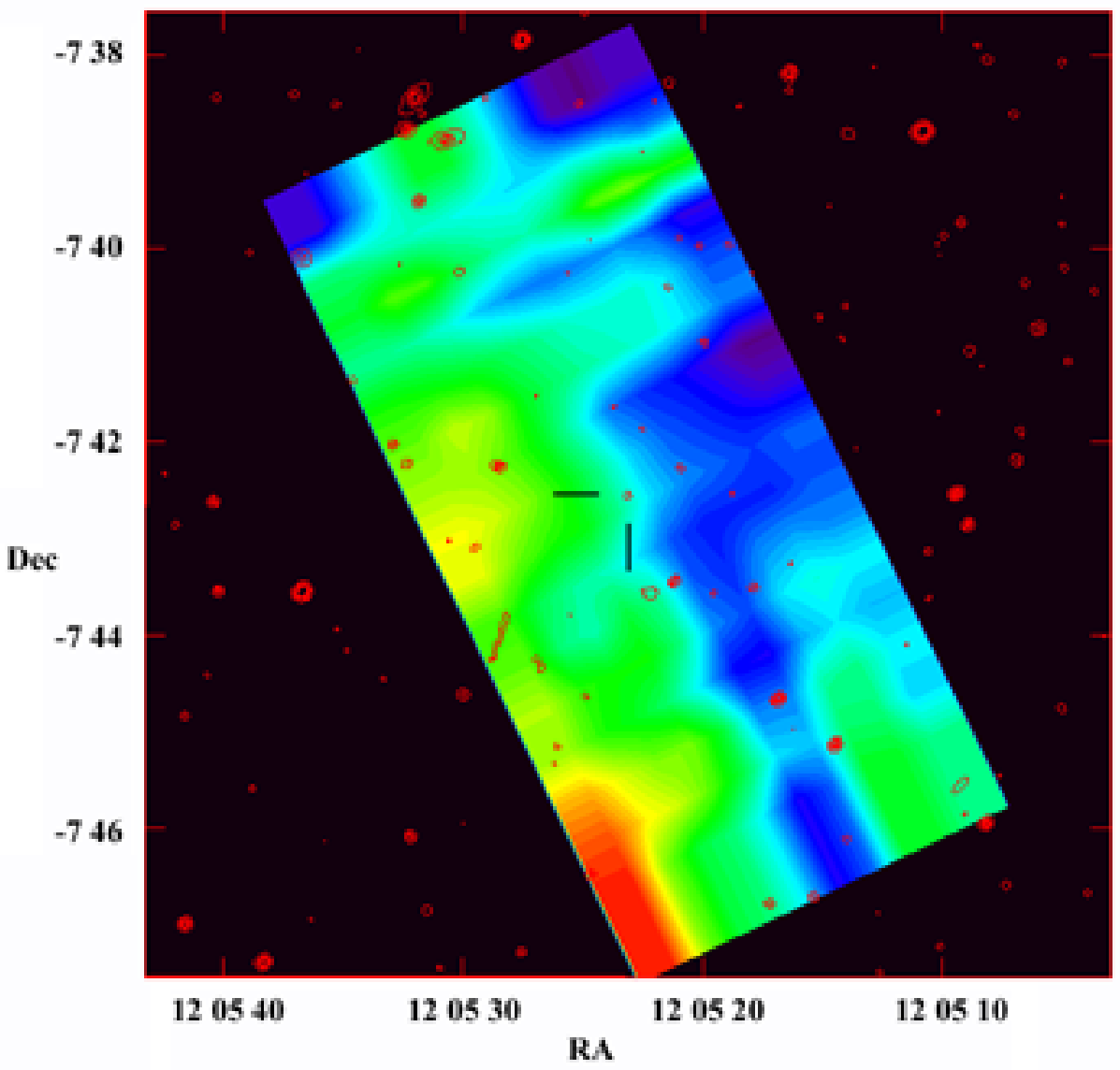}{The 170\ums ISOPHOT map. The quasar is located in the 
centre of the image but has not been detected by ISOPHOT. Its position is
marked. (A colour version of this image is available at the MNRAS website.)}
{fig:phot}

Combined results of this and other papers are presented in Table \ref{KL:tab2}. 
Figure \ref{fig:fluxes} shows the IR data, along with a greybody spectrum at a 
temperature of 68K. This was derived using the method outlined in Benford 
\etals \shortcite{ben99}, i.e.
$$
S_\nu = B_\nu \Omega( 1 - exp( -\tau) ) 
$$
with $\tau = ( \nu / \nu_0)^\beta$, where $\nu_0$ = 2.4 THz, the critical
frequency at which the source becomes optically thin, and taking $\beta$ = 1.5.
The current data does not support the temperature of 50K, with the other
parameters, presented in Benford \etals \shortcite{ben99}. It is, however, 
consistent with the higher temperature of 68K as given in Isaak \etals 
\shortcite{isa94}. The maximum temperature allowed by the 170\ums upper limit 
is 80K, although then the fit to the longer wavelength data becomes poor.

Figure \ref{fig:fluxes2}, shown as normalised $L_\nu$ against $\nu$, extends 
the previous figure to include the 6cm data. It can be compared with the
average spectrum for a Radio-Quiet Quasar given in either Figure 10 of Elvis
\etals \shortcite{elvis94} or Figure 7 of Polletta \etals \shortcite{poll00}. 
If we approximately normalise the fluxes of BR1202-0725 at 6cm and R to those 
of the average RQQ the 15\ums datapoint is also in agreement, whereas the 350 
-- 800 \ums fluxes from BR1202-0725 are higher than the average RQQ, indicating 
the presence of cool dust in this quasar.

\figs{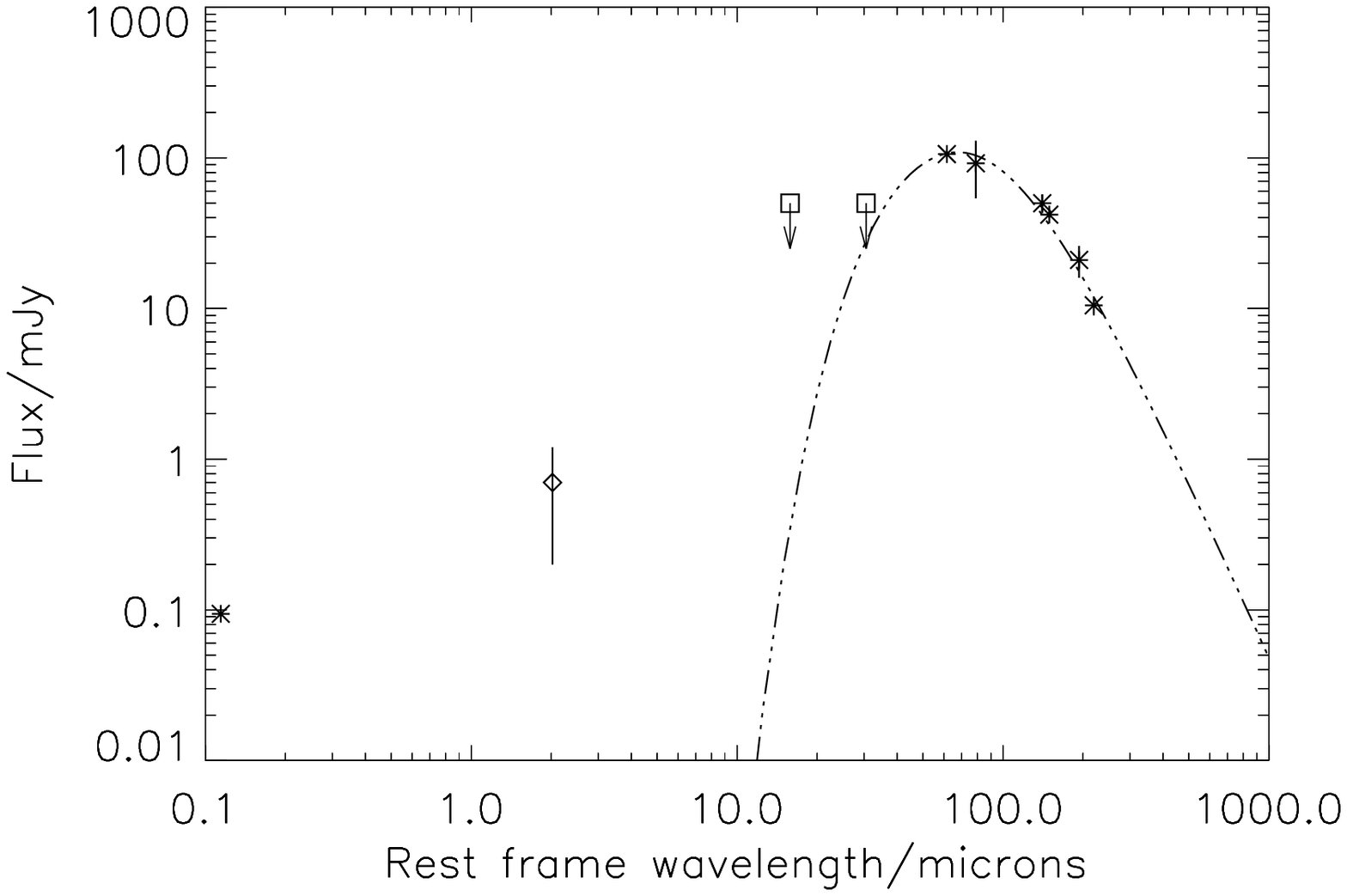}{Data points from Table \ref{KL:tab2} plotted with a 
black-body of temperature 68K. The ISOCAM datapoint is plotted as a diamond, 
the ISOPHOT 3$\sigma$ upper limits are plotted as squares and observations with 
other instruments are plotted as asterisks. The 68K black-body is shown as a 
dash-dot line.}{fig:fluxes}

\figs{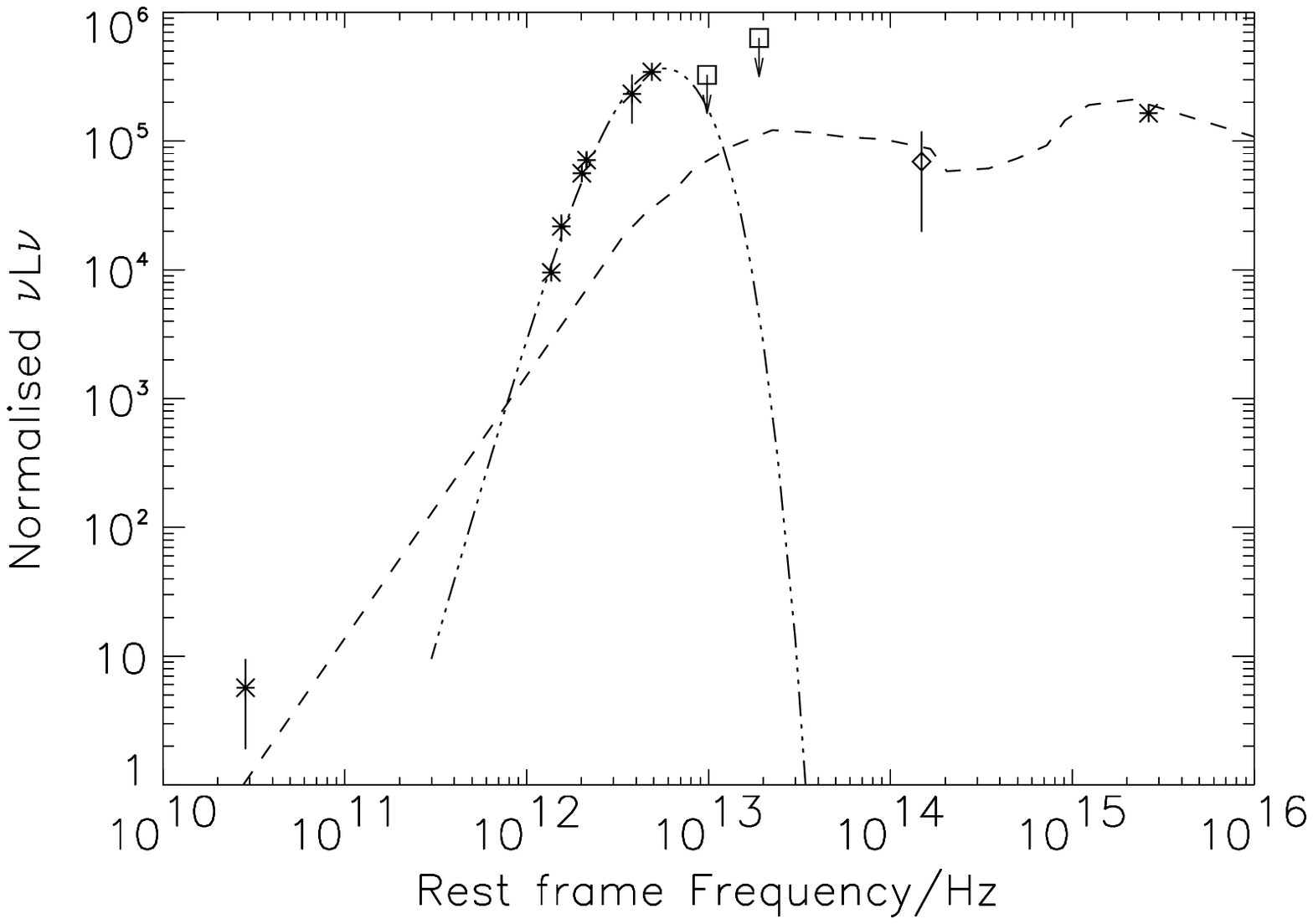}{All data points from Table \ref{KL:tab2} plotted as 
relative luminosity against frequency. The ISOCAM datapoint is plotted as a 
diamond, the ISOPHOT 3$\sigma$ upper limits are plotted as squares and 
observations with other instruments are plotted as asterisks. The 68K 
black-body is shown as a dash-dot line, and the average RQQ SED from Elvis 
\etals \shortcite{elvis94} is shown as a dashed line. The 15\ums flux of 
BR1202-0725 is consistent with the average RQQ, whereas it exhibits an excess 
of emission at the far-IR wavelengths.}{fig:fluxes2}

\begin{table}
 \begin{minipage}{115mm}
 \caption{BR1202-0725 fluxes.}
 \label{KL:tab2}
 \begin{tabular}{@{}llll}
Observed	&Rest		&Flux		&Precision \\
Wavelength	&Wavelength	&		&\, \\
\um		&\um		&mJy		&mJy\\
\hline
0.65\rlap{$^1$}	&0.114		&0.094		& -- \\
11.5		&2.02		&0.7		& 0.2\\
90		&15.8		&$<$ 50$^5$     & \\
170		&30.6		&$<$ 50$^5$     & \\
350\rlap{$^2$}	&61.5		&106		& 7 \\
450\rlap{$^3$}	&79.1		&92		& 38 \\
800\rlap{$^3$}	&141		&50		& 7 \\
1100\rlap{$^3$}	&193		&21		& 5 \\
1250\rlap{$^1$}	&220		&10.5		& 1.5 \\
1250\rlap{$^4$}	&220		&12.59		& 2.28 \\
60000\rlap{$^1$}&10500          &0.3            & 0.2 \\
 \end{tabular}
 \end{minipage}

$1$. McMahon \etals \shortcite{mcm94}.

$2$. Benford \etals \shortcite{ben99}.

$3$. Isaak \etals \shortcite{isa94}.

$4$. Omont \etals \shortcite{omo96a}.

$5$. These are 3$\sigma$ upper limits.
\end{table}

\section{Conclusions}

We have detected the distant, dusty quasar BR1202-0725 in the near-IR and 
obtained upper-limits for the flux at two far-IR wavelengths. The upper-limits 
for the flux levels at 90 and 170\um, when combined with the previous 
ground-based measurements, are consistent with the far-IR and sub-mm being
emitted from a grey-body, most probably arising from dust. The maximum 
temperature allowed by the 170\ums upper limit is 80 K, while the probable
temperature is 68K.

The near-IR flux level of 0.7$\pm$0.2 mJy at 11.5\ums is at a level consistent
with with originating in a normal quasar at the distance of BR1202-0725.

\section*{Acknowledgments}

The authors thank Robert Priddey for a fruitful discussion concerning models of
BR1202-0725.

This paper is based on observations with the Infrared Space Observatory (ISO).
ISO is an ESA project with instruments funded by ESA member states (especially
the PI countries: France, Germany, the Netherlands and the United Kingdom) and
with the participation of ISAS and NASA.

The ISOPHOT data presented in this paper were reduced using PIA, which is a 
joint development by the ESA Astrophysics Division and the ISOPHOT Consortium 
with the collaboration of the Infrared Processing and Analysis Center (IPAC). 
Contributing ISOPHOT Consortium institutes are DIAS, RAL, AIP, MPIK, and MPIA.

The ISOCAM data presented in this paper were analysed using `CIA', a joint 
development by the ESA Astrophysics Division and the ISOCAM Consortium. The 
ISOCAM Consortium is led by the ISOCAM PI, C. Cesarsky.

\end{document}